\documentclass[a4paper,superscriptaddress,aps,prb,twocolumn,floatfix,citeautoscript]{revtex4}
\usepackage{times}
\usepackage{newtxmath}

\newcommand{\vect}[1]{\mathbf{#1}}
\usepackage{graphicx}
\usepackage{color}
\usepackage{ulem}
\usepackage{amsmath}
\usepackage[english]{babel}
\usepackage[utf8]{inputenc}

\usepackage{bm}
\usepackage{url}
\usepackage[colorlinks=true,linkcolor=blue,citecolor=blue,urlcolor=blue]{hyperref}
\graphicspath{{figures/}}

\begin{document}
%
%
\title{Photoemission spectrum in paramagnetic FeO under pressure: towards an \textit{ab initio} description}
\newcommand{\lpt}{Laboratoire de Physique Th\'eorique - IRSAMC, CNRS, Universit\'e Toulouse~III - Paul Sabatier, 118 Route de Narbonne, F-31062 Toulouse, France}
\newcommand{\lcpq}{Laboratoire de Chimie et Physique Quantiques - IRSAMC, CNRS, Universit\'e Toulouse~III - Paul Sabatier, 118 Route de Narbonne, F-31062 Toulouse, France}
\newcommand{\etsf}{European Theoretical Spectroscopy Facility (ETSF)}
\author{S. Di Sabatino}
\affiliation{\lpt}
\affiliation{\lcpq}
\affiliation{\etsf}
\author{J. Koskelo}
\affiliation{\lpt}
\affiliation{\etsf}
\author{J.~A. Berger}
\affiliation{\lcpq}
\affiliation{\etsf}
\author{P. Romaniello}
\email{pina.romaniello@irsamc.ups-tlse.fr}
\affiliation{\lpt}
\affiliation{\etsf}
%

\keywords{...}

\begin{abstract}

In this work we provide an exhaustive study of the photemission spectrum of paramagnetic FeO under pressure using a refined version of our recently derived many-body effective energy theory (MEET). We show that, within a nonmagnetic description of the paramagnetic phase, the MEET gives an overall good description of the photoemission spectrum at ambient pressure as well as the changes it undergoes by increasing pressure. In particular at ambient pressure the band gap opens between the mixed Fe $t_{2g}$ and O $2p$ states and the Fe 4s states and, moreover, a $d$-$d$ gap opens, which is compatible with a high-spin configuration (hence nonzero local magnetic moments as observed in experiment), whereas decreasing pressure the band gap tends to close, $t_{2g}$ states tend to become fully occupied and $e_{g}$ fully unoccupied,  
which is compatible with a low-spin configuration (hence a collapse of the magnetic moments as observed in experiment). This is a remarkable result, since, within a nonmagnetic description of the paramagnetic phase, the MEET is capable to correctly describe the photoemission spectrum and the spin configuration at ambient as well as high pressure.  
For comparison we report the band gap values obtained using density-functional theory with a hybrid functional containing screened exchange (HSE06) and a variant of the $GW$ method (self-consistent COHSEX), which are  reliable for the description of the antiferromagnetic phase. Both methods open a gap at ambient pressure, although, by construction, they give a low-spin configuration; increasing pressure they correctly describes the band gap closing. We also report the photoemission spectrum of the metallic phase obtained with one-shot fully-dynamical $GW$ on top of LDA, which gives a spectrum very similar to DMFT results from literature.
\end{abstract}
\date{\today}
\maketitle
\section{Introduction}
FeO is a basic oxide component of the Earth's interior \cite{MURTHY303} and has a rather complex pressure-temperature phase diagram.\cite{Pasternak_PRL1997,Ozawa_PRB2011,Ohta_PRL2012,Atou_2004,Rueff_2005} It was found experimentally that, under ambient conditions, it is a paramagnetic insulator with a rock-salt B1 crystal structure and it undergoes a phase transition into rhombohedrally distorted B1 structure above 16 GPa.\cite{Akimoto_1985}  At high temperature and pressure an insulator-to-metal transition is observed with a collapse of the local magnetic moment.
\cite{Ohta_PRL2012,Pasternak_PRL1997} A similar magnetic collapse at high pressure is also measured in other similar  transition-metal (TM) oxides such as MnO, CoO, and NiO.\cite{Atou_2004,Yoo_PRL2005,Rueff_2005,Gavriliuk_PRL2012}

Various detailed theoretical studies of the electronic structure and phase stability of these transition metal oxides, in particular in the paramagnetic phase (PM), which employ dynamical mean-field theory (DMFT) combined with \textit{ab initio} band-structure methods, are present in the literature, which interpret these systems as Mott insulators. \cite{Isaak_PRB1993,Cohen_Science1997,Shorikov_PRB2010,Leonov_PRB2015,Leonov_PRB2016} However, Trimarchi \textit{et al.} \cite{ZUNGER} have shown that a correct description of these systems can also be obtained using only band structure theories provided that one models the spin-disordered PM phase using a larger supercell where each TM  site can have different spin direction and different local bonding geometry, but still with a zero total spin. Within this so-called ``polymorphous" band structure description the band gap opening is not driven by strong correlation as in DMFT. 

In this work we investigate the electronic structure of paramagnetic FeO both at ambient and high pressure 
within an \textit{ab initio} framework. We adopt the DMFT standpoint, in which the existence of local magnetic moments is a consequence of the electron localization and not an essential part of the gap opening mechanism itself. We therefore model the system as nonmagnetic (spin-unpolarized). 
Paramagnetic FeO is a special case because within a nonmagnetic modelling of the PM phase, Hartree-Fock (HF) theory is enough to open a gap in this system, unlike in similar systems, such as MnO, CoO, and NiO.\cite{stefano_JCTC} This occurs because non-local exchange well separates in energy the transition metal $d$ states with $t_{2g}$ symmetry from those with $e_{g}$ symmetry. Since Fe has six $d$ electrons, they fully occupy the $t_{2g}$, opening a gap. In other similar transition-metal oxides, this does not happen since the transition metal has less than or more than six electrons, and therefore they partially occupy the $t_{2g}$ or $e_g$ bands, leading to a metal.  This happens because in the nonmagnetic case static mean-field theories can only open a gap if there is an even number of electrons, in which case a band insulator forms, or if there is a spontaneously broken spin and/or translational symmetry (e.g., magnetic order). Of course within such a nonmagnetic description of the paramagnetic phase local magnetic moments are zero by construction. 
Mean-field methods are more appropriate in the high-pressure regime, where the band gap closes  and the system is in a low-spin configuration. We show, for example, that generalized Kohn-Sham theory within the HSE06 functional \cite{HSE06} can describe the band-gap closing by increasing pressure.
Instead, using a refined version of our recently derived many-body effective energy theory (MEET), one can get an overall good description of the photoemission spectrum and spin configuration at ambient and high pressure, yet remaining in a nonmagnetic description of the PM phase. 
This paper is organized as follows: in Sec.~\ref{Theory} we describe the MEET; in Sec.~\ref{Comp_details} we summarize the computational details of the performed calculations. Results of the photoemission spectrum at ambient as well as high pressure are presented and discussed in Sec.~\ref{Results}.
Summary and conclusions are drawn in Sec.~\ref{Conclusions}.

\section{Theoretical background: the MEET\label{Theory}}

Within the many-body effective energy theory (MEET) the time-ordered 1-body Green's function $G(\omega)$, which gives the spectral function as $A(\omega)=1/\pi |\text{Im}G(\omega)|$, is split into removal ($R$) and addition ($A$) parts as $G(\omega) = G^{R} (\omega) + G^{A} (\omega)$. The spectral function is the key quantity which is related to photoemission spectra. The diagonal matrix elements of $G^{R/A}(\omega)$ can be written in terms of an effective energy $\delta_{i}^{R/A} (\omega)$ as:
\begin{eqnarray}
 G_{ii}^{R} (\omega)  &=& \frac{n_i}{\omega-\delta_i^{R}(\omega)-\text{i}\eta}, \\
 G_{ii}^{A} (\omega)  &=& \frac{1-n_i}{\omega-\delta_i^{A}(\omega)+\text{i}\eta},
 \end{eqnarray} 
where we use the basis of natural orbitals, i.e., the orbitals which diagonalize the one-body reduced density matrix (1-RDM), with $n_i$ the occupation number of state $i$. 
Since the spectral function is expressed as
\begin{equation}
A_i(\omega)=n_i\delta(\omega-\delta^{R}_i(\omega))+(1-n_i)\delta(\omega-\delta^{A}_i(\omega))
\label{Eqn:SF_MEET}
\end{equation}
fractional occupation numbers could lead to a gap opening in the degenerate $d$ states. We will come back to this point in  Sec.~\ref{Sec:MEET}.
The effective energy $\delta_{i}^{R/A} (\omega)$ can be written as an expansion in terms of reduced density matrices. The expression truncated at the level of the one- and two-body reduced density matrices (2-RDM) reads
\begin{eqnarray}
  \delta_i^{R,(1)} &=& h_{ii}+\sum_jV_{ijij}n_j+\frac{1}{n_i}\sum_jV_{ijkl}
  \Gamma^{(2)}_{\text{xc},klji}
  \label{Eqn:MEET_removal}
  \\
\delta_i^{A,(1)} &=& h_{ii}+\sum_jV_{ijij}n_j\nonumber\\
& &-\frac{1}{(1-n_i)}\sum_j\left[V_{ijji}n_j+V_{ijkl}
\Gamma^{(2)}_{\text{xc},klji}
\right],
 \label{Eqn:MEET_addition}
\end{eqnarray}
where $h_{ij}=\int d\mathbf{x}\phi^*_i(\mathbf{x})h(\mathbf{r})\phi_j(\mathbf{x})$ are the matrix elements of the one-particle noninteracting Hamiltonian $h(\mathbf{r})=-\nabla^2/2+v_{\text{ext}}(\mathbf{r})$ and $V_{ijkl}=\int d\mathbf{x}d\mathbf{x}'\phi^*_i(\mathbf{x})\phi^*_j(\mathbf{x}')v_c(\mathbf{r},\mathbf{r}')\phi_k(\mathbf{x})\phi_l(\mathbf{x}')$ are the matrix
elements of the Coulomb interaction $v_c$. Here we approximate the exchange-correlation contribution to the 2-RDM in terms of the power functional
$\Gamma^{(2)}_{\text{xc},klji}=-n_i^{\alpha}n_j^{\alpha}\delta_{il}\delta_{jk}$, with $0.5\le \alpha\le 1$.  Moreover the natural orbitals and occupation numbers are obtained within reduced-density matrix functional theory (RDMFT),\cite{Gilbert75} where again the  power functional \cite{sharma_PRB08} 
is employed to approximate the 2-RDM in the total energy minimization. The approximation to the 2-RDM is extremely important, because the performance of the MEET heavily relies on it. 
The power functional has been tested on various extended systems,\cite{sharma_PRB08,PhysRevA.79.040501,PhysRevLett.110.116403} but the results on the occupation numbers are scarce.\cite{PhysRevB.75.195120}
In the MEET one needs accurate occupation numbers and this is not clearly guaranteed using the power functional. At high pressure, indeed, we have to vary $\alpha$ in order to have a low-spin configuration. This is discussed in the next section.

The MEET, at the level of $\delta^{R/A,(1)}$, gives qualitatively good spectra in gapped materials, but hugely overestimates the band gap; moreover it opens a gap even in metals. More details can be found in Refs \onlinecite{stefano,stefano_JCTC}. This is mainly due to the cut-off of the MEET equations at the level of the 1- and 2-RDMs. This can be understood by reformulating $\delta^{R/A}_i(\omega)$ in terms of moments $\mu^{R/A}_{n,i}=\sum_k B_{ii}^{k,R/A}(\epsilon^{R/A}_k)^n/\sum_k B_{ii}^{k,R/A}$  (with $n$ indicating the order of the moment and $B_{ii}^{k,R/A}$ the spectral weight of the energy $\epsilon_k^{R/A}$) of $G^{R/A}_{ii}(\omega)$, as reported in Ref.~\onlinecite{stefano_JCTC}. The approximation $\delta^{R/A,(1)}_i$ is equal to the first moment $\mu_{1,i}^{R/A}$, which is a weighted average of all of the poles of $G^{R/A}_{ii}(\omega)$. Let us suppose that the photoemission spectrum is composed of quasiparticle peaks with roughly 30$\%$ \cite{lucia_book} 
of the spectral weight transferred to the corresponding satellites for the weakly correlated states (i.e., states characterized by natural occupation numbers close to 1 or 0) and a larger percentage for the strongly correlated (states characterized by fractional natural occupation numbers) $d$ states. Let us also assume that there is a single dominant plasmon which generates the satellites. The approximation $\delta^{R,(1)}_i$ ($\delta^{A,(1)}_i$) produces a removal (addition) peak which is red shifted (blue shifted) with respect to the exact quasiparticle; the shift is proportional to the spectral weight of the corresponding satellites. This explains: i) the  overall good relative position of the spectral features in the removal and the addition parts of the MEET spectrum, with the $d$ states slightly red (blue) shifted with respect to rest of the removal (addition) part; ii) the overestimation of the band gap.
An approximate estimation of the influence of the 3-RDM on the spectrum by screening the interaction in front of the 2-RDM in Eq.~\eqref{Eqn:MEET_removal} indicates an improvement of the band gap.\cite{stefano} We notice that also the approximations to the 2-RDM play a role in the final result.
Work to understand the impact of the two approximations and how to include the effect of higher-order RDMs is currently in progress. 
Here we overcome the band-gap problem of the MEET using a pragmatic approach  by aligning the weakly correlated MEET states (namely O $2p$ and Fe $4s$ for the removal and the addition parts, respectively) to the LDA states. The rationale 
behind this procedure is the following: i) removal and addition parts of the PES, separately, are overall well described by the MEET and a rigid shift of the two parts towards one another would suffice to reduce the fundamental gap; ii) in the spirit of DMFT the weakly correlated states are assumed to be well described by LDA. The alignment yields the rigid shift we are seeking. In the following we will refer to this protocol as MEET+LDA.

\begin{figure*}[t]
\begin{center}
\includegraphics[width=0.47\textwidth]{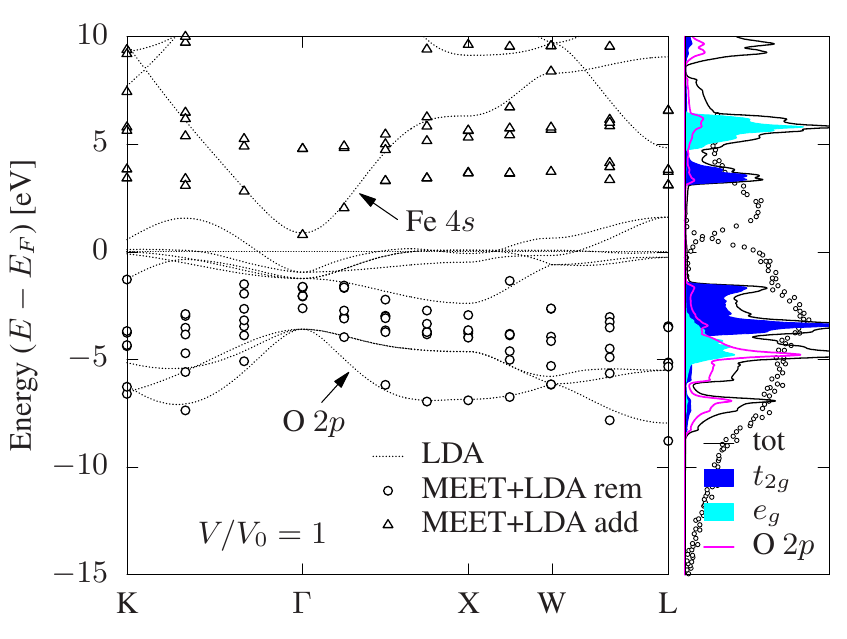}
\hspace{15pt}
\includegraphics[width=0.47\textwidth]{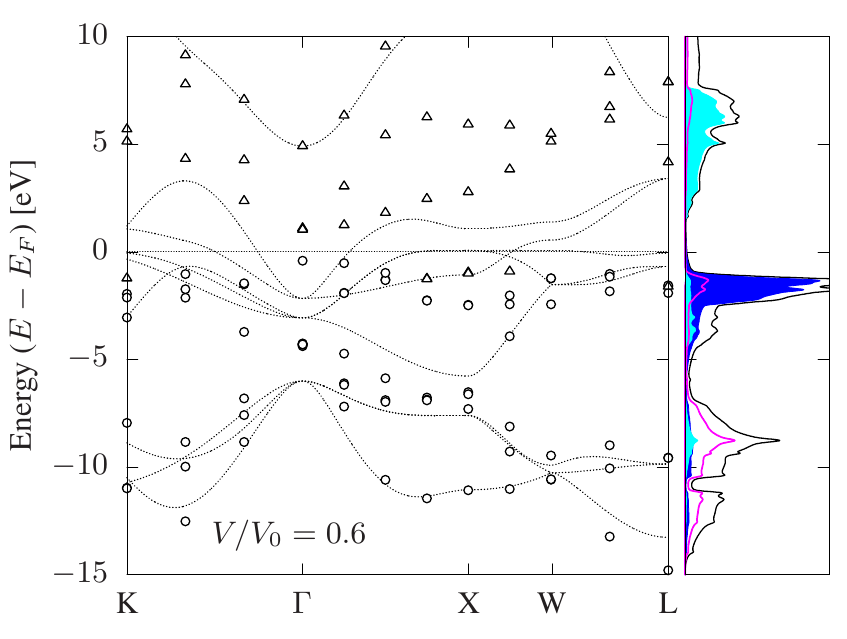}
\caption{MEET+LDA $\mathbf{k}$-resolved and total spectral function. Removal (circles) and addition energies (triangles)  are reported along high symmetry directions for relative volumes $V/V_0=1$ (left panel) and $V/V_0=0.6$ (right panel), where $V_0$ is the equilibrium volume at ambient pressure.  We also report the LDA band structure (black dotted line).
The contributions of the Fe $t_{2g}$ and $e_g$ states and of the O $2p$ states to the total spectral function are also reported.
The experimental photoemission spectrum is taken from Ref.~\onlinecite{Zimmermann_1999}.
}
\label{Fig:MEETalignedLDA}
\end{center}
\end{figure*}

\section{Computational details\label{Comp_details}}
The MEET calculations have been performed using a modified version of the open-source full-potential linearized augmented plane wave (FP-LAPW) code \textsc{Elk},\cite{elk2} with practical details of the calculations following the scheme described in Ref.~\citenum{sharma_PRB08}. 
HSE06 and self-consistent Coulomb hole plus screened exchange (scCOHSEX) calculations have been performed using the Vienna Ab-initio Simulation Package (\textsc{Vasp})\cite{Kresse.Furthmuller:1996,Shishkin2006_PRB,Kresse_PRL07,paier2006} with the projector-augmented wave (PAW) method.\cite{Blochl,Kresse.Joubert:1999:PRB} For calculations of $G_0W_0$ spectral function we used the \textsc{Abinit} code \cite{gonze2005,gonze2009} employing planewaves and norm-conserving pseudopotentials. 

We described the PM phase of FeO as nonmagnetic and we used a rocksalt structure with the experimental lattice constant, i.e.,  4.33 \AA.\cite{Hentschel:1970}

For the \textsc{Elk} calculations we used a  $\Gamma$-centered $8\times8\times8$ $\vect{k}$-point grid.
The muffin-tin radius for Fe and O atoms are 2.31 and 1.73 a.u., respectively.
The muffin-tin (MT) radius times maximum $|\vect{G}+\vect{k}|$ vectors, $R^{\text{MT}}\times\text{max}{|\vect{G}+\vect{k}|}$, is 7.0, while the maximum length of $|\vect{G}|$ for expanding the interstitial density and potential is 12.0 $\mbox{a.u.}^{-1}$. We included 8 empty bands.

In the $G_0W_0$ and scCOHSEX calculations we treated the dielectric screening at the level of the random-phase approximation (RPA), and included Fe $3s$, $3p$, $3d$, and $4s$ and O $2s$ and $2p$ as valence states. In the \textsc{Vasp} calculations we included 200 bands for screening and self-energy, and expanded the pseudowavefunctions and dielectric matrices in the basis of plane waves up to 700 eV and 800 eV, respectively. We used $\Gamma$-centered $8\times8\times8$ $\vect{k}$-point grid. 

In the \textsc{Abinit} calculations we employed optimized norm-conserving Vanderbilt pseudopotentials,\cite{hamann2013,vansetten2018} and expanded the ground-state wavefunctions up to 90 Ha. We used 100 bands for building the screening and self-energy. In the screening calculations we expanded the wavefunctions and screening matrix in plane waves up to 30 and 10 Ha, respectively. In the self-energy calculations we used planewaves up to 60 Ha for both wavefunctions and Fock operator. We employed $\Gamma$-centered $12\times12\times12$ $\vect{k}$-point grid. The frequency dependence of the self-energy was treated with the contour-deformation technique.\cite{lebegue2003} We sampled the imaginary axis with 25 frequencies, and real axis with 40 frequencies up to 20 eV.  

\begin{figure}[b]
\begin{center}
\includegraphics[width=0.47\textwidth]{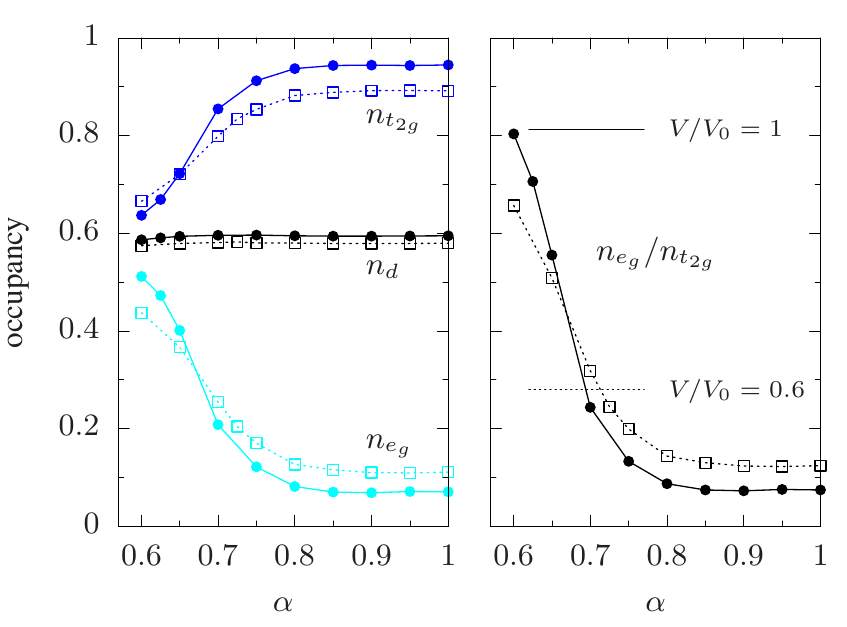}
\caption{(Left panel) Occupancies of $t_{2g}$ (blue) and $e_g$ (cyan) orbitals at relative volume $V/V_0=1$ (solid line and filled circles) and $V/V_0=0.6$ (dashed line and open squares) as a function of the power functional parameter $\alpha$. We report also the average $d$ occupancy (black). 
In the right panel the ratio between $e_g$ and $t_{2g}$ occupancies is plotted. Values of the ratio $n_{eg}/n_{t2g}$ are extracted from Ref.~\onlinecite{Leonov_PRB2015} and reported as horizontal lines.
}
\label{Fig:varalpha}
\end{center}
\end{figure}

\begin{figure*}[t]
\begin{center}
\includegraphics[width=0.42\textwidth]{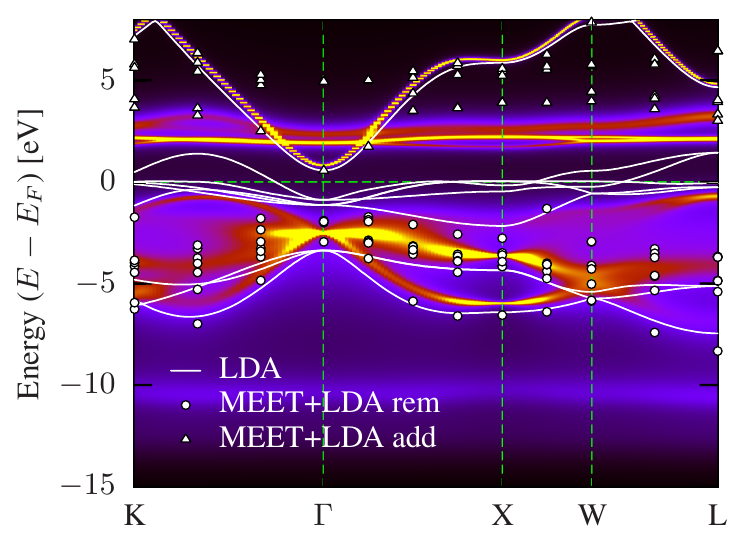}\hspace{25pt}
\includegraphics[width=0.42\textwidth]{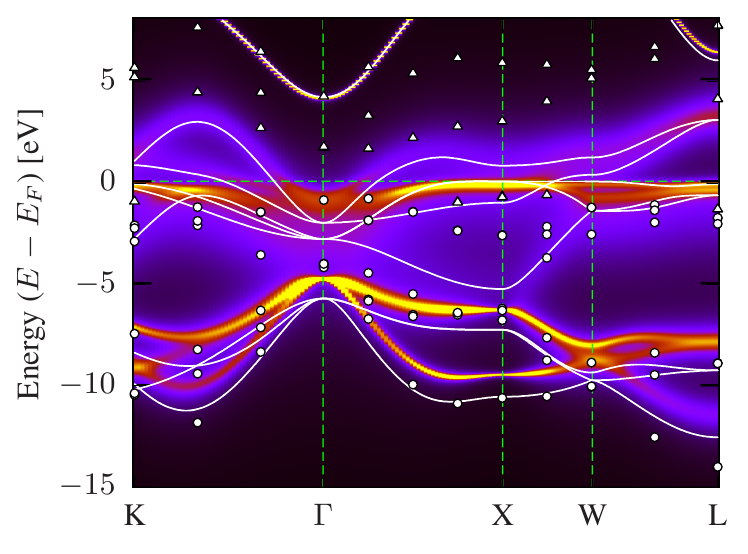}
\caption{$\mathbf{k}$-resolved spectral function of FeO: MEET+LDA energies are reported along high symmetry directions and compared with DMFT results from Ref.~\onlinecite{Leonov_PRB2015} for ambient  (left) and high pressure (right). We also report LDA band structures (white solid line) which are used to align the MEET energies. Note that for a meaningful comparison we have used
for MEET and LDA calculations the same lattice constant used in  Ref.~\onlinecite{Leonov_PRB2015}.}
\label{Fig:DMFT}
\end{center}
\end{figure*}

\section{Results\label{Results}: Spectral function}
\subsection{Many-body Effective Energy Theory\label{Sec:MEET}}
In Fig.~\ref{Fig:MEETalignedLDA} we report the spectral function of PM FeO at ambient pressure (left panel) and at high pressure (right panel) calculated using the MEET+LDA.
The MEET removal and addition energies are aligned using the LDA bands corresponding to the weakly-correlated O $2p$ and Fe $4s$ states respectively.

At ambient pressure we use $\alpha=0.65$ for the power functional, as suggested in literature.\cite{sharma_PRB08} At high pressure we use $\alpha=0.70$. This value guarantees a decrease of the $e_g$ occupancy in favour of the $t_{2g}$ occupancy, as illustrated in Fig.~\ref{Fig:varalpha}, and thus a transition to a low-spin configuration.
More specifically this value of $\alpha$ is fixed by the $t_{2g}/e_g$ occupation ratio at relative volume $V/V_0=0.6$ (with $V_0$ the equilibrium volume at ambient pressure) being approximately equal 
to the value deduced from Ref.~\onlinecite{Leonov_PRB2015}.\footnote{
We notice that the equilibrium volume used in Ref.~\onlinecite{Leonov_PRB2015} is the result of a geometry optimisation within DMFT, whereas we use the experimental value. Therefore in order to compare our results with the results of Ref.~\onlinecite{Leonov_PRB2015} we use an interpolation of the data reported in Ref.~\onlinecite{Leonov_PRB2015}.
}

Unfortunately, there are very few experiments reporting the band gaps of FeO in the PM phase at ambient pressure, because of the difficulty to prepare pure FeO samples.  We only found the value of 2.4 eV from an optical absorption measurement.\cite{BOWEN1975355} 
Overall the theoretical photoemission spectrum compares well with the experimental one.\cite{Zimmermann_1999}
The calculated band gap opens between the mixed Fe $t_{2g}$ and O $2p$ states and the Fe $4s$ states with a value of about 2.1 eV. The $d$-$d$ gap is instead about 4.4 eV. The MEET+LDA $\mathbf{k}$-resolved spectral function compares rather well with DMFT results from Ref.~\onlinecite{Leonov_PRB2015} as one can see from Fig.~\ref{Fig:DMFT}. 

The gap opening in the $d$-state manifold occurs because for values of $\alpha<1$ the occupation numbers, which enter the expression of the MEET spectral function [see Eq.~\eqref{Eqn:SF_MEET}], are fractional and hence, each of them gives rise to a removal energy and an addition energy. This is not the case for $\alpha=1$ (which is Hartree-Fock), since in this case the occupation numbers are either 1 or 0, and therefore the $t_{2g}$ states, which are fully occupied, will give rise only to removal energies, and the $e_g$ states, which are fully unoccupied, will give rise only to addition energies. 
The Fe $t_{2g}/e_g$ orbital contributions to the spectral function reported in Fig.~\ref{Fig:MEETalignedLDA} are compatible with a high-spin configuration and, therefore, in line with the experimental observation of a local magnetic moment. This is a quite remarkable result, since, within a nonmagnetic description of the PM phase, the MEET+LDA is capable to give an overall good description of both the spectral function and the  high-spin configuration. 

\begin{figure*}[t]
\begin{center}
\includegraphics[width=0.40\textwidth]{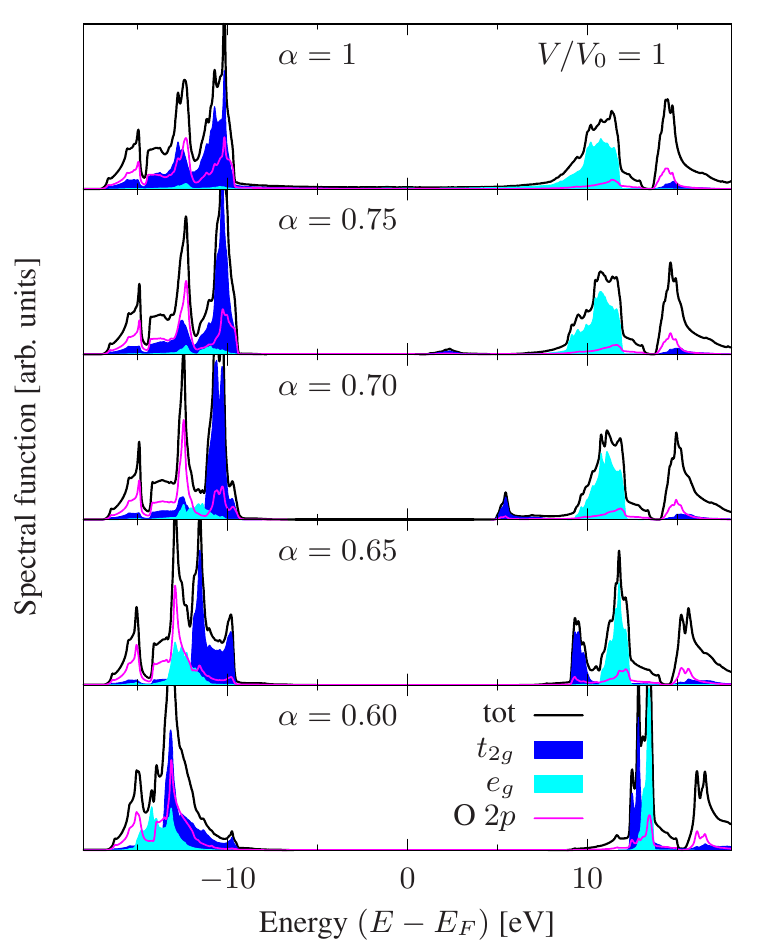}\hspace{20pt}
\includegraphics[width=0.40\textwidth]{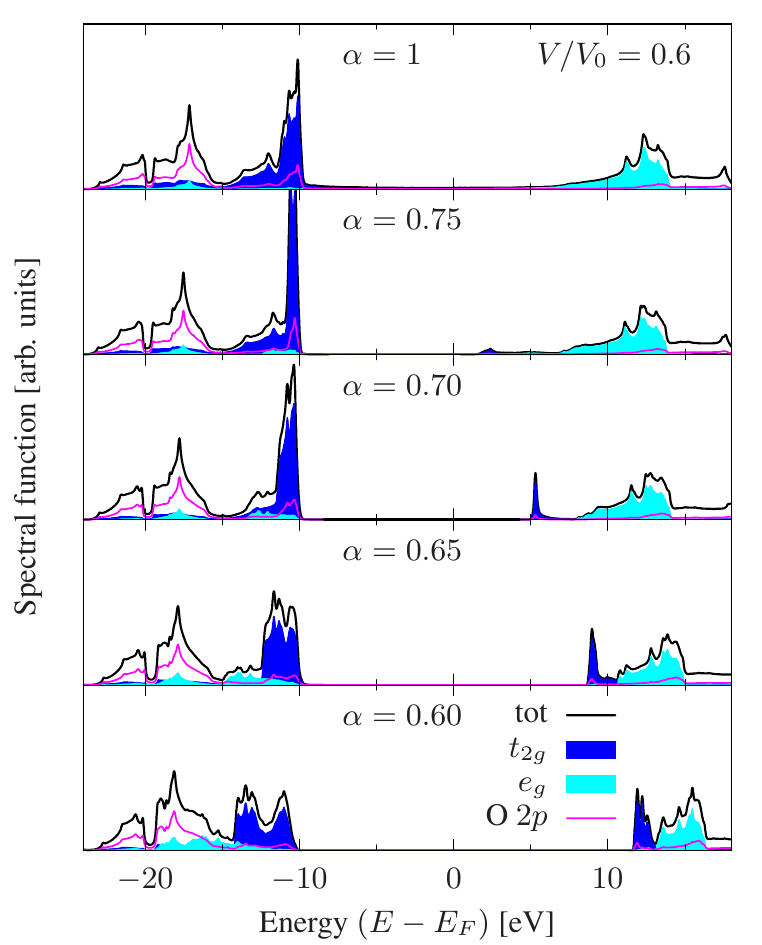}
\caption{Photoemission spectrum calculated using MEET (i.e., without the alignment with LDA) for various values of $\alpha$ for $V/V_0=1$ (left panel) and $V/V_0=0.6$ (right panel). The contributions of the Fe $t_{2g}$ and $e_g$ states and of the O $2p$ states to the total DOS are also reported.}
\label{Fig:MEET_2}
\end{center}
\end{figure*}

In Fig.~\ref{Fig:MEET_2} we also report the spectral function of FeO at ambient pressure and high pressure (right panel) calculated using the MEET (i.e., without the alignment) with various values of the $\alpha$ parameter. We see that increasing $\alpha$, and hence decreasing the degree of correlation treated, tends to reduce the gap but also to deform the spectrum, in particular because the $t_{2g}$ component in the addition part of the spectrum starts to migrate to lower energies while decreasing in intensity, as the occupation numbers tend towards 0 or 1. As already anticipated, at $\alpha=1$ one gets the HF spectrum, which is compatible with a low-spin configuration. We also notice that HF opens a gap also at high pressure. 

In the right panel of  Fig.~\ref{Fig:MEETalignedLDA} 
we report the spectral function of FeO at high pressure (at relative volume $V/V_0=0.6$).
At this reduced volume MEET+LDA results go in the right direction: the band gap closes, although the spectral weight around the Fermi energy is still too  small (at least with respect to DMFT results, see Ref.~\onlinecite{Leonov_PRB2015}). The MEET+LDA $\mathbf{k}$-resolved spectral function compares well with DMFT results from Ref.~\onlinecite{Leonov_PRB2015}, although the $d$ conduction bands are largely blue shifted with respect to DMFT results, as one can see from Fig.~\ref{Fig:DMFT}.
We notice that to compare our results with DMFT we use the same lattice constant as in  Ref.~\onlinecite{Leonov_PRB2015}, namely 4.42~\AA \, (8.36 a.u.), for the MEET+LDA results reported in Fig.~\ref{Fig:DMFT}.  

\begin{table} [b]
\caption{Fundamental band gap (eV) vs. relative volume $V/V_0$. The experimental value at ambient pressure is 2.4 eV from an  optical  absorption measurement\cite{BOWEN1975355}}
\begin{ruledtabular}
\begin{tabular}{lcc}
$V/V_0$ & HSE06 & scCOHSEX \\
\hline
1     &  2.9  &  4.1  \\
0.75  &  2.2  &  3.0  \\
0.625 &  1.5  &  1.9 \\
0.50   &  0.4  &  0 \\
0.45   &  0  &  0 \\
\end{tabular}
\end{ruledtabular}
\label{bandgap}
\end{table}
\begin{figure}[b]
\begin{center}
\includegraphics[width=0.47\textwidth]{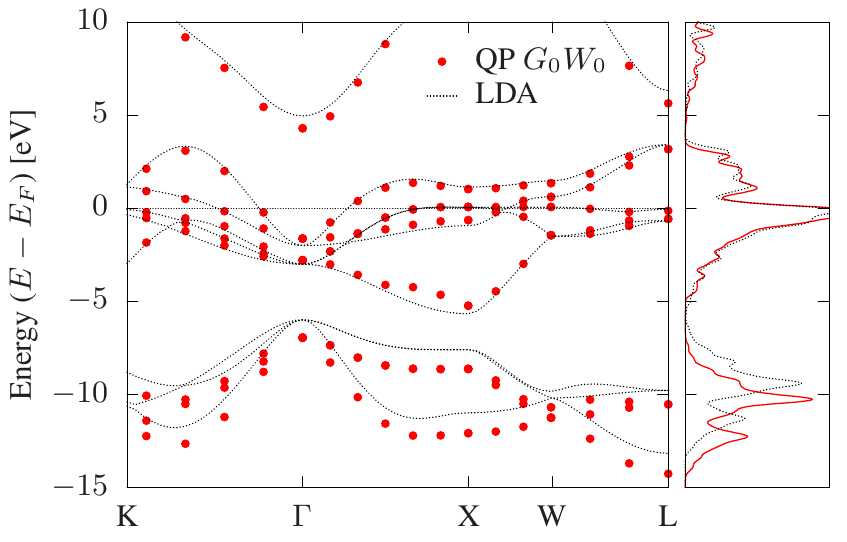}
\caption{Quasiparticle band structure (red dots) and spectral function (red solid line) within $G_0W_0$ at $V/V_0=0.6$. For comparison we report also the LDA band structure and DOS (dotted black line).}
\label{Fig:QP_GW}
\end{center}
\end{figure}

\begin{figure}[t]
\begin{center}
\includegraphics[width=0.42\textwidth]{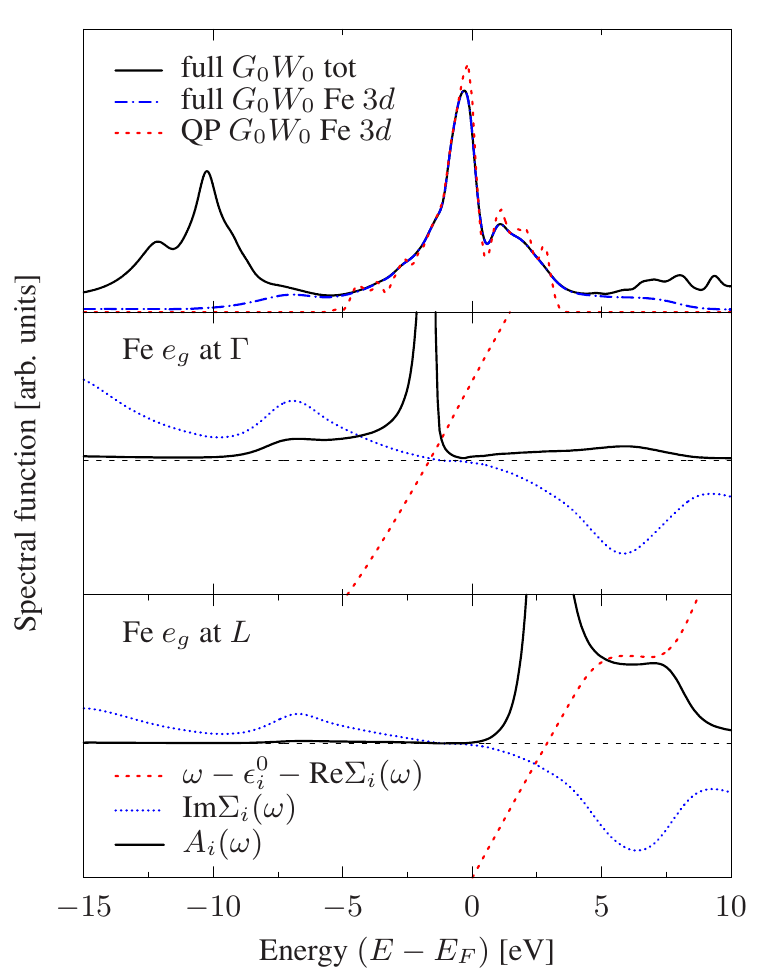}
\caption{(Upper panel) Spectral function calculated with the full dynamical $G_0W_0$ (black solid line). For Fe $3d$ states we report also the
 partial spectral function (blue dot-dashed line)  
and the QP spectral function (red dashed line). We notice that, for comparison, the QP spectrum has been corrected introducing the corresponding renormalization factor $Z_i$ which on average is 0.64 for the Fe-$3d$ bands.
(Middle and bottom panels)
Spectral function $A_i(\omega)$ (solid black line), $\omega-\epsilon_i^0-\text{Re}\Sigma_i(\omega)$ (red dashed line), and $\text{Im}\Sigma_i(\omega)$ (blue dotted line) for the Fe $e_g$ states at the $\Gamma$ and $L$ points. The quasiparticle energy $\epsilon^{\text{QP}}_i$ is given by the solution of the equation $\omega-\epsilon^0_i-\text{Re}\Sigma_i(\omega)=0$.}
\label{Fig:full_GW}
\end{center}
\end{figure}

\subsection{HSE06 and $GW$}
For comparison we also report our results obtained within the hybrid functional HSE06 and $GW$, which have been applied to transition metal oxides in the AF phase at ambient pressure with relatively good results \cite{Rodl,Faleev_PRL2004,Efstratios_PRB2015}. 
In Table \ref{bandgap} we report the band gap of FeO vs. the relative volume $V/V_0$ calculated using HSE06 and scCOHSEX. COHSEX is a static variant of $GW$ \cite{Hedin65}, which allows for a computationally more feasible self-consistent solution; being static it reproduces only the quasiparticle peaks in the spectral function. While these methods give an incorrect picture at ambient pressure, as already discussed in the introduction, they are justified at high pressure, where the fundamental band gap closes and the system goes in a low-spin configuration. We observe that within both methods the system is a metal at $V/V_0=0.45$, with scCOHSEX closing the gap slightly faster (with respect to the volume decrease) than HSE06. 
In Fig.~\ref{Fig:QP_GW} we report the quasiparticle (QP) band structure and DOS of the metallic phase (at $V/V_0=0.6$) calculated with one-shot $GW$ ($G_0W_0$) on top of LDA (using the linearized QP equation). For comparison also the LDA band structure and DOS are reported. $GW$ induces a renormalization (although small) of the $d$ bands  and lower the O $2p$ bands, with an increase of the gap with the $d$ states. The full (i.e., taking into account the full frequency dependence of the self-energy) $G_0W_0$ spectral function reported in Fig.~\ref{Fig:full_GW} (upper panel) is in overall good agreement with DMFT results from literature, although the description of satellites could improve by introducing some form of self-consistency.\cite{gatti2013}
The shoulder at about -7 eV and the peak around 7 eV largely come from $d$-band satellites as one can see in Fig.~\ref{Fig:full_GW} (upper panel), where the QP and full $G_0W_0$ spectral functions are compared. These satellites are missing in the MEET within the current static approximation to $\delta_i^{R/A}(\omega)$. By analyzing the satellites of the $e_g$ states at the $\Gamma$ (for occupied $e_g$ states) and $L$ (for unoccupied $e_g$ states) points (see middle and lower panel of Fig.~\ref{Fig:full_GW}), we identify the same satellites as in the total spectral function in the upper panel of Fig.~\ref{Fig:full_GW}. A similar scenario occurs for the $t_{2g}$ bands. This analysis is based on the expression of the total spectral function $A(\omega)=\sum_{i}A_i(\omega)$ with 
 \begin{equation}
 A_i(\omega)=\frac{1}{\pi}\frac{|\text{Im}\Sigma_i(\omega)|}{|\omega-\epsilon_i^0-\text{Re}\Sigma_i(\omega)|^2+|\text{Im}\Sigma_i(\omega)|^2},
 \end{equation}
 where $\epsilon_i^0$ are the noninteracting single particle energies, and $\text{Im}\Sigma_i$ and $\text{Re}\Sigma_i$ are the imaginary and real parts, respectively, of the self-energy.
 Both satellites are located  in the vicinity of structures in
the imaginary part of the $GW$  self-energy, which we checked to arise from a plasmon peak at $\approx$5 eV in the energy loss function.   

\section{Summary and Conclusions\label{Conclusions}}
 We presented a detailed \textit{ab initio} study of the photoemission spectrum (PES) of paramagnetic FeO at ambient and high pressures using a refined version of the many-body effective energy theory (MEET). Our protocol is based on three important steps: i) treat removal and addition parts of the PES separately; ii) assuming the relative position of the MEET spectral features correct in both parts; iii) aligning the MEET energies corresponding to weakly correlated states with the LDA ones in order to get a rigid shift which decreases the band gap. We showed that this protocol  gives an overall good description of the spectral function and of the spin configuration both at ambient and high pressure even modelling the PM phase as nonmagnetic. This is a very important result of our method, contrary to state-of-the-art \textit{ab initio} methods which do not correctly describe the system at ambient pressure. However, more advanced approximations to the 1- and 2-RDMs are needed in order to make our method fully predictive. Work in this direction is in progress. For completeness we also report the results obtained using density-functional theory with the  hybrid functional HSE06 and some variants of the $GW$ method (namely, scCOHSEX and $G_0W_0$ on top of LDA). Whereas the two methods are not appropriate at ambient pressure, increasing pressure they correctly describe the band gap closing. The $G_0W_0$ photoemission spectrum of the metallic phase is in overall good agreement with DMFT results from literature. Besides the $d$-states quasiparticle peaks around the Fermi energy, two satellite peaks are also present, both stemming from plasmonic structures in the imaginary part of the $G_0W_0$ self-energy. These peaks are not described by the MEET within the current static approximation to the effective energy $\delta^{R/A}_i(\omega)$.
 
\begin{acknowledgments}
This study has been supported through the EUR grant NanoX no.~ANR-17-EURE-0009 in the framework of the ``Programme des Investissements d’Avenir" and by ANR (project no.~ANR-18-CE30-0025, no.~ANR-19-CE30-0011). The authors would like to thank Matteo Gatti and Claudia R\"odl for fruitful discussions.
\end{acknowledgments}

\end{document}